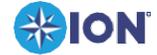

REGULAR PAPERS

# Single-Satellite Navigation on Lunar North Pole

Tim Gong[1] | Andrew Dempster*[2]

[1]School of Electrical Engineering and Telecommunications, University of New South Wales, Sydney, NSW, Australia

[2]Australian Center for Space Engineering Research, University of New South Wales, Sydney, NSW, Australia

**Correspondence**

*Andrew Dempster, a.dempster@unsw.edu.au

**Abstract**

The Moon is a primary focus of space exploration. Current navigation methods face significant limitations in providing precise location data for lunar missions. In particular, existing methods often require direct Line of Sight to Earth, have limited capacity, and suffer from long signal travel times. This paper aims to tackle these challenges through a novel single satellite navigation system at the lunar North Pole. By utilising the Doppler effect, this system facilitates 3D geolocation of a stationary receiver on the lunar surface. Key findings include choosing a Low Lunar Orbit (LLO) suitable for North Pole coverage, designing a 3-step geolocation algorithm tailored to lunar conditions, constructing a comprehensive error budget, and evaluating the system performance through Dilution of Position (DOP).

## 1 | INTRODUCTION

In recent years, there has been continued interest in space missions, particularly lunar missions. In early April 2024, the White House has asked NASA to create a Coordinated Lunar Time (LTC) for the Moon by 2026 (Ramirez-Simon, 2024). This suggests numerous lunar missions in the near future. The success of these missions relies on infrastructure not just a timing system but also the accurate positioning of operating equipment.

Current navigation techniques, which depend on direct communication with Earth, face significant limitations due to the need for a direct line of sight, limited capacity, and long signal travel times (NASA, 2024). These constraints are particularly challenging for operations in the lunar polar regions where direct LoS is not always feasible. To address these challenges, this work proposes a single satellite lunar navigation system specifically for the North Pole region. Specifically, the receiver will utilise the Doppler frequency in the received signal from the satellite to determine a point solution in 3D space.

This work builds upon Psiaki (2021), which details single satellite navigation for Earth-based applications using LEO satellites; and Coimbra et al. (2024), which details using an Elliptic Lunar Frozen Orbit satellite for Lunar Pathfinder. This work is novel in that it is the first to use Low Lunar Orbit as the navigation satellite, the first not to require any prior knowledge of the receiver location, and the first to consider the mirror location involved in Doppler methods.

The primary purpose of this lunar navigation system is to fill the gap until a comprehensive lunar global navigation system is established. Currently, it can assist in various aspects of lunar exploration, including facilitating more effective lunar surface exploration, increasing resource mapping accuracy, and setting the possibility of lunar-based infrastructure development. In particular, water-ice deposits have been found in some permanently shadows areas in the North Pole region (Spudis et al., 2010). This lunar navigation system can aid the extraction of the precious water resources, which is crucial for long-term exploratory missions and even human habitation.

The remainder of the paper is organised as follows: Section 2 proposes an LLO suitable for lunar north pole navigation. Section 5 details the error budget for a Doppler navigation system on the Moon. Section 3 explains the generation of Doppler measurement data. Section 4 outlines the algorithm to map the Doppler measurement to a point in space. Section 7 shows a metric for the expected location accuracy. Section 6 shows the results from simulation. Finally, Section 8 concludes the paper.

## 2 | LUNAR ORBITS

The Low Lunar Orbit (LLO) from (Lara et al., 2009) is used as the proposed orbit in this paper. It has an average altitude of around 120 km, resulting in a period of 2 hours. It is a polar orbit, providing maximum coverage for the poles. The orbital parameters are shown in Table 1. It was simulated in STK and shown to be stable for at least 10 years.

| Parameter | Symbol | Value |
| --- | --- | --- |
| Semi-major axis | $a$ | 1860.52 km |
| Eccentricity | $e$ | 0.0359457 |
| Inclination | $i$ | 90 deg |
| Argument of perilune | $\omega$ | 270 deg |
| RAAN | $\Omega$ | 0 deg |

**TABLE 1**
Initial orbital parameters of proposed orbit (Lara et al., 2009)

The LRO orbit from 2009 to 2016 can be estimated using method explained in (Mazarico et al., 2018). The data are archived at the annex MIT LOLA PDS Data Node https://imbrium.mit.edu/LRORS. The two orbit paths are shown in Figure 1. From visual inspection, the paths of the orbits almost overlap with each other. This supports the claim of the stability of the proposed orbit.

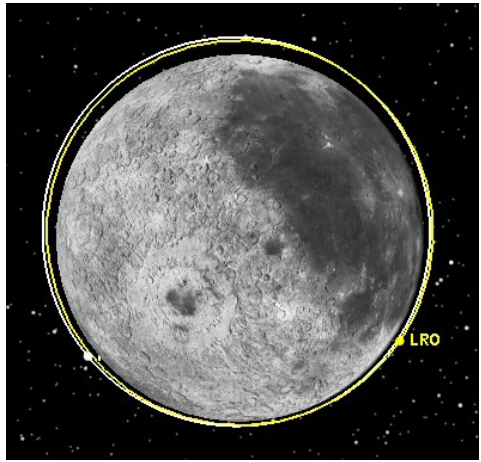

**FIGURE 1** Proposed orbit (white) and LRO (yellow)

## 3 | DOPPLER DATA GENERATION

Tables 2, 3 and 4 explain the symbols used in this work. The measured carrier Doppler shift can be written as (Psiaki, 2021)

$$D = -\frac{1}{\lambda_0} \frac{d\Delta\rho_{ADR}}{dt_R}.$$

TABLE 2
Constants

| Symbol | Value | Unit | Meaning |
|---|---|---|---|
| $c$ | 299792.458 | km/s | Speed of light |
| $R_M$ | 1737.4 | km | Radius of Moon |
| $f_0$ | 2050 | MHz | Nominal transmission frequency |
| $\lambda_0$ | $c/f_0$ | km | Nominal transmission wavelength |

TABLE 3
Symbols

| Symbol | Meaning |
|---|---|
| $j$ | Each observation |
| $N$ | Total number of observations |
| $D^j$ | The $j$th measurement of the carrier Doppler shift |
| $\Delta\rho^j_{ADR}$ | The $j$th accumulated delta range |
| $t_R$ | Received time of each measurement |
| $\sigma^j_{tot}$ | Total measurement error of the $j$th measurement |
| $\rho^{\hat{}j}$ | Normalised displacement from satellite to receiver |

**Measured Frequency** The noisy measured Doppler frequency is generated as (Psiaki, 2021)

$$D^j = -\frac{1}{\lambda_0} \frac{-(\hat{\rho}^j)^T v^j \left( \frac{1 + \frac{d\delta^j}{dT^j}}{1 + \frac{1}{c} \frac{a^j - (\rho^{\hat{}j})^T v^j}{\delta t_p}} \right) + c\frac{d\delta_R}{dT_R} - c\frac{d\delta^j}{dT^j}}{1 + \frac{d\delta_R}{dT_R}}. \tag{1}$$

The propagation delay is given by

$$\delta t^j_p = \frac{1}{c} \| r^j_{sr}(r; t_R) \|.$$

$\omega_M = \frac{2\pi}{27.321661 \times 86400}$ rad s$^{-1}$ is the Moon's rotation rate since each sidereal month is 27.321661 days (Erik Gregersen, 2024). The direction cosines matrix

$$A\left(\omega_M \delta t^j_p\right) = \begin{pmatrix} \cos\left(\omega_M \delta t^j_p\right) & \sin\left(\omega_M \delta t^j_p\right) & 0 \\ \omega_M \delta t^j_p & -\cos\left(\omega_M \delta t^j_p\right) & 0 \\ 0 & 0 & 1 \end{pmatrix}$$

compensates for the rotation of the lunar coordinate frame while the signal travels from the satellite to the receiver. The correction term for the time rate of change of Moon's rotation is

$$a^j_{\delta t_p} = -\langle \hat{\rho}^j, \begin{bmatrix} 0 & 0 & \omega_M \end{bmatrix}^T \times A\left(\omega_M \delta t^j_p\right) r^j_s\left(t_R - \delta^j_r - \delta t^j_p\right) \rangle,$$

where $\langle \cdot, \cdot \rangle$ gives a dot product. The satellite velocity is given by

$$v^j = A\left(\omega_M \delta t^j_p\right) \frac{dr^j_{sr}}{dt} \Big|_{(t_R - \delta t^j_p)}.$$

**Calculated Frequency** It can be seen that $-\lambda_0 D^j$ is the measured relative speed between satellite and receiver, and $\frac{d\Delta\rho^j_{ADR}}{dt^R}$ is the $j$th calculated relative speed using the ephemeris data. The accumulated delta range can be modelled as

$$\Delta\rho^j_{ADR}(r; t_R) = \| r^j_{sr} \| + c(\delta^j_r - \delta^j_s), \tag{2}$$

where $\delta^j_r$ and $\delta^j_s$ are the receiver and satellite clock error, and

$$r^j_{sr}(r; t_R) = r - A\left(\omega_M \delta t^j_p\right) r^j_s\left(t_R - \delta^j_r - \delta t^j_p\right)$$

is the position vector from satellite at time of transmission to the unknown receiver location $r$ at time of reception.

The derivative of $\Delta\rho^j_{ADR}$ is computed using a five point difference approximation as

$$\frac{d\Delta\rho^j_{ADR}}{dt_R}(r;t_R) \approx \frac{1}{12\Delta t_R}[\Delta\rho^j_{ADR}(r;(t_R - 2\Delta t_R)) \\ -8\Delta\rho^j_{ADR}(r;(t_R - \Delta t_R)) \\ +8\Delta\rho^j_{ADR}(r;(t_R + \Delta t_R)) \\ -\Delta\rho^j_{ADR}(r;(t_R + 2\Delta t_R))], \quad (3)$$

where $\Delta t_R$ is the nominal finite difference interval of $t_R$. This work uses $\Delta t_R = 0.1$s as suggested by Psiaki (2021).

## 4 | LOCATION ALGORITHM

A 3-step algorithm is used to map the measurements to a location. Unlike current methods, no prior knowledge of the receiver location is required. Constants and symbols used throughout these steps are shown in Table 2 and 3.

### 4.1 | Step 1 - Algebraic Solution

Nguyen and Dogancay (2016) gives the procedure to find a rough initial estimate of the receiver location through algebraic methods. For each satellite pass, the satellite locations in the middle 100 seconds are used. This ensures that enough data is used for the algorithm, and the data with the highest C/N0 is used.

Figure 2 gives an illustration of how this method works. It starts by assuming constant speed of the satellite $V$ and altitude $h$.

The first step is to estimate the point of closest approach $\hat{x}_{P\ CA}$. This is the satellite position with the closest distance to the receiver. This can be calculated as

$$\tilde{\alpha}_j = \cos^{-1}\left(\frac{c}{V}\left(\frac{D^j}{f_0} - 1\right)\right) \\ \hat{x}_{P\ CA} = \frac{1}{N-1}\sum_{j=2}^{N}\left(\frac{(x_j - x_1)\sin\tilde{\alpha}_1 \cos\tilde{\alpha}_j}{\sin(\tilde{\alpha}_j - \tilde{\alpha}_1)} + x_j\right). \quad (4)$$

The second step is to estimate the receiver-subtrack distance, which is given as

$$\hat{d} = \sqrt{\frac{\sum_{j=1}^{N}(\hat{x}_{P\ CA} - x_j)^2}{\sum_{j=1}^{N}\cot^2\tilde{\alpha}_j} - h^2}. \quad (5)$$

Figure 2 relates the receiver location to $\hat{x}_{P\ CA}, h, \hat{d}$. From this, the receiver location can be found as

$$d_{sat} = x_N - x_1 \\ d_d = d_{sat} \times \hat{x}_{P\ CA} \\ r = \hat{x}_{PCA}\frac{\|\hat{x}_{P,CA}\| - h}{\|\hat{x}_{P\ CA}\|} + \hat{d}\frac{d_d}{\|d_d\|} \\ r_{norm} = \frac{r}{\|r\|}R_M,$$

where $d_{sat}, d_d$ are the direction vectors of the satellite subtrack and $d, r$ is the receiver location. $r_{norm}$ is projection of $r$ onto lunar surface to increase estimation accuracy, which is passed to the second step in section 4.2.

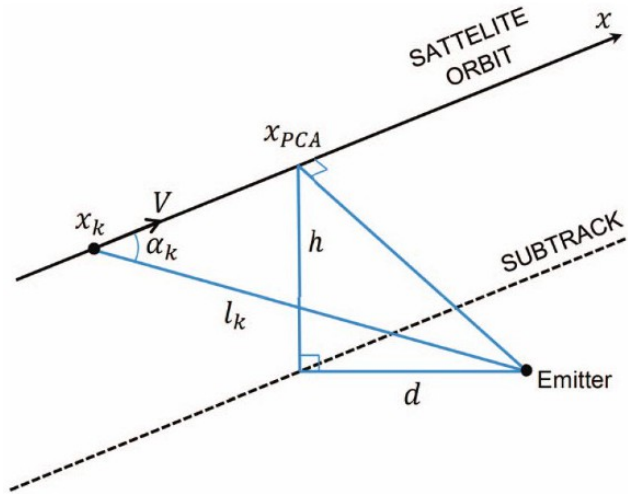

**FIGURE 2** Illustration of satellite and receiver location (shown as Emitter) (Nguyen & Dogancay, 2016)

## 4.2 | Step 2 - Constrained Nonlinear Least Square

This step aims to increase the estimation accuracy to somewhere near the actual location. This step and the next step improve upon Psiaki (2021) and Baron et al. (2024), significantly increasing the convergence rate. The constraint in this step is that the receiver is situated on the lunar surface. This significantly increases convergence rate, as altitude is found to be one of the most dominant variables that affects the cost function values. Algorithm 1 shows the process, while symbols used are summarised in Table 4.

**TABLE 4**
Symbols for section 4.2

| Symbol | Meaning |
| --- | --- |
| $r$ | Receiver position |
| $r_s$ | Satellite position |
| $v$ | Satellite velocity |
| $J$ | Cost function |
| $\delta t_p$ | Propagation delay between satellite to receiver |
| $A(\cdot)$ | Moon rotation matrix |
| $r_{rs}$ | Relative position from satellite to receiver |
| $R$ | Unit vector of $r_{rs}$ |
| $h$ | Calculated velocity |
| $f$ | Residue of measured and calculated velocity |
| $H$ | Derivative of $R$ |

The algorithm uses the Gauss-Newton method to find the solution that minimises the cost function

$$J = \frac{1}{2} \sum_{j=1}^{N} \left( \lambda_0 D^j + \frac{d\Delta\rho^j_{ADR}}{dt_R} \right)^2, \qquad (6)$$

with the constraint that the length of the solution is equal to the radius of Moon. The measurement error $\sigma^j_{tot}$ is assumed to be constant and thus omitted in this section. The threshold in the last line of Algorithm 1 determines the length of the final step, and is set to 1 m as this step does not require high accuracy. The derivative of the unit vector of geometric range $R$ between the

AUTHOR ONE et al---

**Algorithm 1** Constrained Gauss-Newton

**Require:** $r, D, t_R, TLE$
1: **repeat**
2:     **for all** observed $D^j$ **do**
3:         $h^j \leftarrow \frac{d\Delta\rho^j_{ADR}}{dt_R}(r)$      ▷ See Equation 2,3
4:         $f^j \leftarrow \lambda_0 D^j + h^j$
5:         Compute $\dot{R}^j$      ▷ See Equation 7
6:     **end for**
7:     $H \leftarrow \begin{bmatrix} \dot{R}^{1^T} & \dot{R}^{2^T} & \cdots & \dot{R}^{N^T} \end{bmatrix}^T$
8:     $H_{xy} = H(:, 1:2),\ f_{xy} = f(1:2)$      ▷ Consider x,y coordinates only
9:     $\Delta r_{xy} \leftarrow (H_{xy}^T H_{xy})^{-1} H_{xy}^T f_{xy}$
10:     Refine $\Delta r_{xy}$      ▷ See Algorithm 2
11:     $r_{xy} \leftarrow r_{xy} + \Delta r_{xy}$
12:     $r \leftarrow \begin{bmatrix} r_{xy}^T & \sqrt{R_M^2 - \|r_{xy}\|^2} \end{bmatrix}^T$      ▷ Map x,y to x,y,z
13: **until** $\|\Delta r_{xy}\| <$ threshold

---

satellite and receiver can be written as

$$\dot{R}^j = \frac{-1}{\|r_{rs}^j\|}\left(v^j - \langle v^j, r_{rs}^j\rangle \frac{r_{rs}^j}{\|r_{rs}^j\|^2}\right). \tag{7}$$

$$r_{rs}^j = -r_{sr}^j(r; t_R) = r - A\left(\omega_M \delta t_p^j\right) r_s^j\left(t_R - \delta_r^j - \delta t_p^j\right)$$

**Armijo's Rule**

Algorithm 2 outlines the Armijo's rule, a method for efficient gradient descent (Katselis, 2019). Armijo's rule ensures that the cost function decreases sufficiently at each iteration. It does this by checking whether the cost function will decrease at least by a fraction of the expected decrease based on a linear assumption. If not, then the step size is reduced. By Taylor's series, this condition will always be satisfied for a small enough step size.

Three user-defined parameters are required: $\alpha, \epsilon, \beta$.

1. $\alpha$ determines how much the cost function should decrease. Since the main goal of this algorithm is to ensure the cost function will not diverge, this is set to a low value of 0.1.
2. $\epsilon$ is the scaling of the step size and is initially set to 1.
3. $\beta$ is the relative size of the new step if the Armijo's condition is not met. It is set to 0.5 to avoid excessively small steps.

---

**Algorithm 2** Refine $r$ by Armijo's Rule

**Require:** $r, \Delta r$
1: $\alpha \leftarrow 0.1,\ \epsilon \leftarrow 1,\ \beta \leftarrow 0.5$
2: $J_k \leftarrow J(r),\ J_{k+1} \leftarrow J(r + \Delta r)$
3: $\nabla r \leftarrow H^T f$
4: **while** $J_{k+1} - J_k \geq \alpha \epsilon \nabla r \Delta r$ **do**
5:     $\epsilon \leftarrow \beta \epsilon$
6:     $J_{k+1} \leftarrow J(r + \epsilon \Delta r)$
7: **end while**
8: $\Delta r \leftarrow \epsilon \Delta r$

## 4.3 | Step 3 - Unconstrained Nonlinear Least Square

This step is aimed to improve upon the estimation in Step 2 and achieve the highest accuracy. Algorithm 3 outlines the process of Step 3. There are three main differences compared to Step 2:

1. The constraint that the receiver is fixed on the lunar surface is now removed.
2. The cost function includes the total measurement error in each measurement:

$$J_\sigma = \frac{1}{2} \sum_{j=1}^{N} \left( \frac{\lambda_0 D^j + \frac{d\Delta\rho^j_{ADR}}{dt_R}}{\lambda_0 \sigma^j_{tot}} \right)^2, \tag{8}$$

3. Each step is refined by soft line search, outlined in Algorithm 4. This is stricter than Armijo's rule as it has an additional condition that the step is large enough to ensure the step leaves the starting tangent of the initial location (Frandsen et al., 2004). This method ensures that each step is nearly optimised in reducing the cost function.

During each iteration, the weighting vector $w$ is calculated as

$$w = 1/\sqrt{\sigma_{tot}} \tag{9}$$

where $\sigma_{tot}$ is calculated from Equation 14. Two new functions $\varphi$ and $\lambda$ are defined as (Frandsen et al., 2004)

$$\varphi(\alpha) = J(r + \alpha \Delta r) \tag{10}$$
$$\varphi'(\alpha) = \Delta r^T J'(r + \alpha \Delta r) \tag{11}$$
$$\lambda(\alpha) = \varphi(0) + 0.001 \alpha \varphi'(0) \tag{12}$$

---

**Algorithm 3** Unconstrained Gauss-Newton

**Require:** $r, D, t_R$, TLE
1: **repeat**
2:   **for all** observed $D^j$ **do**
3:     $h^j \leftarrow \frac{d\Delta\rho^j_{ADR}}{dt_R}(r)$     ▷ See Equation 2,3
4:     $f^j \leftarrow \lambda_0 D^j + h^j$
5:     Compute $\dot{R}^j$     ▷ See Equation 7
6:   **end for**
7:   $H \leftarrow \begin{bmatrix} {\dot{R}^1}^T & {\dot{R}^2}^T & \cdots & {\dot{R}^N}^T \end{bmatrix}^T$
8:   Compute $w$     ▷ See Equation 9
9:   $f_\sigma \leftarrow \langle w, f \rangle$, $H_\sigma \leftarrow diag(w)H$
10:   $\Delta r \leftarrow \left(H_\sigma^T H_\sigma\right)^{-1} H_\sigma^T f_\sigma^j$
11:   Refine $\Delta r$     ▷ See Algorithm 4
12:   $r \leftarrow r + \Delta r$
13: **until** $\|\Delta r\| <$ threshold

---

**Summary**     There are several reasons for using a three-step process:

1. The first step ensures that the initial estimate for the gradient descent in step 2 is within a reasonable distance to the actual receiver location.
2. The second step uses constrained 2D gradient descent with Armijo's rule to enhance the convergence rate of the traditional GN method. Empirically this step can converge when the initial guess is 300 km away from the actual position. If the GN method is applied without altitude restriction, the estimate can go beyond the lunar surface and into outer space. If Armijo's rule is not used, the estimate can oscillate between several points and take too long to reach the lowest cost function value.
3. The third step allows full 3D gradient descent with weightings corresponding to the total measurement error. This is included because some data points have less variance than others due to received $C/N_0$. This step employs the soft line search method, which is computationally more expensive than Armijo's rule but has better convergence rate (Frandsen et al., 2004).

**Algorithm 4** Refine $r$ by Soft Line Search

```
 1:  α_max ← 10,  k_max ← 100                                    ▷ Maximum step length and iteration
 2:  if φ'(0) ≥ 0 then                                            ▷ See Equation 11
 3:      α ← 0
 4:  else
 5:      k ← 0,  γ ← β · φ'(0)
 6:      a ← 0,  b ← min {1, α_max}
 7:      while φ(b) ≤ λ(b) and φ'(b) ≤ γ and b < α_max and k < k_max do    ▷ See Equation 10
 8:          k ← k + 1,  a ← b
 9:          b ← min {2b, α_max}
10:      end while
11:      α ← b
12:      while (φ(α) > λ(α) or φ'(α) < γ) and k < k_max do        ▷ See Equation 12
13:          k ← k + 1
14:          Refine α and [a, b]                                   ▷ See Algorithm 5
15:      end while
16:      if φ(α) ≥ φ(0) then
17:          α ← 0
18:      end if
19:  end if
20:  end
```

**Algorithm 5** Refine $\alpha$ and $[a, b]$ in soft line search

```
 1:  begin
 2:  D := b − a,  c := (φ(b) − φ(a) − D · φ'(a))/D²
 3:  if c > 0 then
 4:      α := a − φ'(a)/(2c)
 5:      α := min {max {α, a + 0.1D}, b − 0.1D}
 6:  else
 7:      α := (a + b)/2
 8:  end if
 9:  if φ(α) < λ(α) then
10:      a := α
11:  else
12:      b := α
13:  end if
14:  end
```

## 4.4 | Mirror Location

The above steps utilise only the measured Doppler frequencies to determine a location estimate. When only one satellite pass is available, two minima of the cost function will be present. One is the actual location, the other is the mirror location, which is symmetrical about the satellite subtrack. For Earth-based applications, Argentiero and Marint (1979) proves a method that maximises the probability of choosing the correct location, which simplifies to comparing the cost function values when no systematic error is present in the system. It claims that the method can be effective > 99% of the time. However, as the Moon self rotation speed is 30 times less than the Earth, this method does not give the desired performance. An alternative method is to increase the observation time to cover at least two satellite passes. Due to the slight difference in the orientation of the two satellite subtracks, the two mirror locations will be at different locations, while the correct location will be at the same place. The slow rotation of the Moon will cause a few km of difference in the mirror locations, which is enough to be distinguished from the actual location.

## 5 | ERROR BUDGET

Several sources of error that affect the location accuracy of the proposed method have been identified. Considering the error budget for GPS, which contains 6 items: ephemeris, receiver clock, receiver measurement noise, ionosphere, troposphere, and multipath. For lunar applications, there is no ionosphere and troposphere; the effect of multipath is minimal due to the lack of lunar infrastructure. However, the satellite clock error needs to be considered as lunar orbiters do not have clocks with the same accuracy as GPS satellites. This results in 4 categories of errors: ephemeris, satellite clock, receiver clock, and measurement noise.

### 5.1 | Ephemeris

Two methods of ephemeris modelling are considered: one takes into account the difficulty in modelling LLO satellites, the other assumes a more ideal situation where we can predict LLO satellites with high accuracy. The two methods result in the upper and lower bounds of the system performance.

#### 5.1.1 | Ephemeris Method 1

This method considers the worst case scenario, and it is achievable with currently available technologies. Three steps are involved: observed orbit, predicted orbit, and fitted orbit. Table 5 shows the error introduced in each step.

**TABLE 5**
Error in ephemeris modelling in each stage in metres, in rms ± std

|  | Along-track | Cross-track | Radial | Total (x,y,z) |
|---|---|---|---|---|
| Observed (Mazarico et al., 2018) | $5.29 \pm 1.11$ | $4.05 \pm 1.03$ | $0.27 \pm 0.06$ | $6.70 \pm 1.47$ |
| Predicted (Granier et al., 2024) | $49.14 \pm 32.39$ | $9.28 \pm 9.27$ | $1.87 \pm 1.61$ | $51.29 \pm 51.17$ |
| Fitted |  |  |  | $0.67 \pm 0.67$ |

**Observed Orbit**   The orbit first has to be observed. The observation error of the LRO is used here due to the similarity between the LRO orbit and the proposed orbit. Table 5 records the error for altitudes between 70 and 90 deg (Mazarico et al., 2018). The estimation period is on average 2.5 days, which contains 3 complete observations of White Sands passes, where the White Sands station in New Mexico is specifically built for LRO, and tracks LRO over 8-10 h daily. The daily orbit of LRO is available from NAIF (Calk et al., 2024).

**Predicted Orbit**   HALO, a lunar propagator, is used to simulate the satellite path over one day (Granier et al., 2024). A one-day period is chosen because the observed orbits are released daily (Calk et al., 2024). 100 simulations are performed with random epoch times between 16 Mar 2024 and 14 Jun 2024 where no manoeuvre activities are performed by the LRO. The actual location of LRO is obtained from Calk et al. (2024), and passed into the propagator to get the predicted orbit. The prediction error is the difference between the actual orbit and the predicted orbit. The rms and std of these errors are shown in Table 5.

Another popular propagator is STK. Figure 3 shows a comparison of the prediction error of HALO and STK in 14 hours. It can be seen that HALO has a smaller error in both position and velocity, giving a more accurate prediction than STK. However, both propagator errors are not white noise, but follows the orbital frequency. To generate a realistic error at a sampling time of 1 s, coloured noise is generated through a second order IIR filter

$$H(z) = \frac{1 - z^{-2}}{1 - 1.9999z^{-1} + 0.9999z^{-2}}. \tag{13}$$

The frequency response of the filter is compared against the DFT of the error in Figure 4. The coloured noise is generated by applying the filter on white noise, then normalised to have the desired standard deviation as shown in Table 5.

**Fitted Orbit**   Cortinovis et al. (2023) suggests that for LLO, ephemeris models based on Chebyshev polynomials are better suited than classical Keplerian parameters due to the actual orbit deviating a lot from the ideal shape as a result from uneven lunar gravity field. In this paper, a 10-order Chebyshev polynomial is fitted on the position of the satellite. Empirically this adds 0.67 m

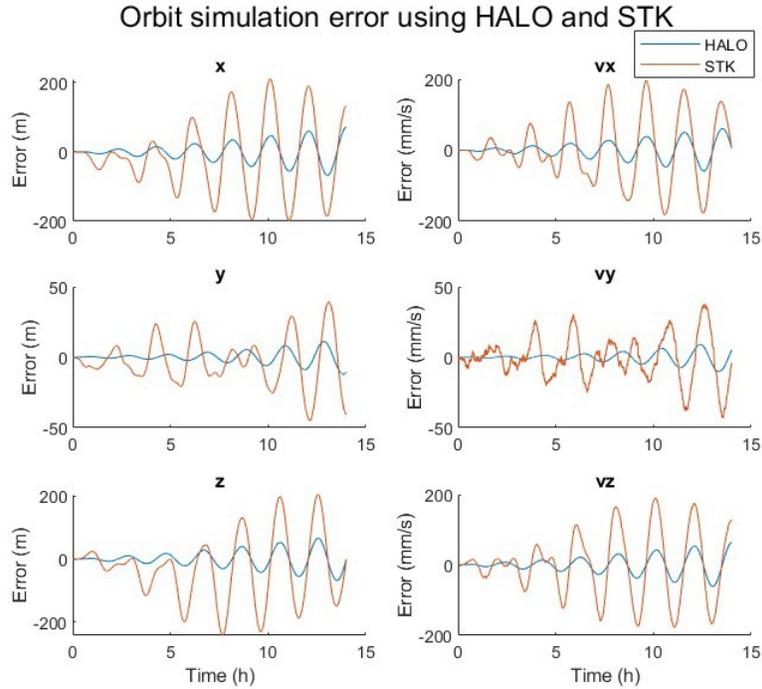

**FIGURE 3** Simulation error using HALO (Granier et al., 2024) and STK

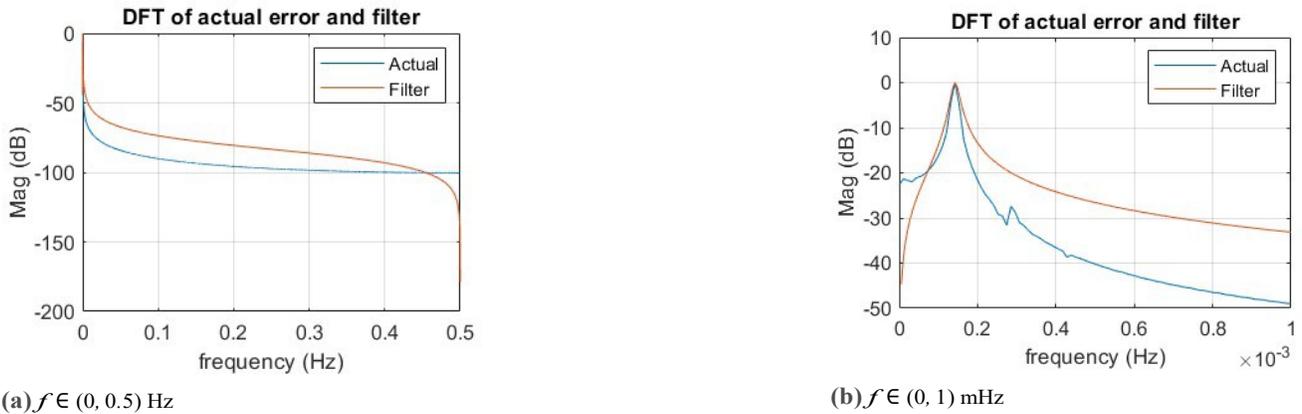

(a) $f \in (0, 0.5)$ Hz

(b) $f \in (0, 1)$ mHz

**FIGURE 4** DFT of actual error (normalised) and frequency response of IIR filter (normalised). (b) is a zoom-in of (a).

to the standard deviation of the ephemeris error. The velocity can be approximated as the derivative of the position (Montenbruck & Gill, 2000).

**Ephemeris Generation Procedure**

1. The proposed orbit is propagated using HALO. This is assumed to be the "true" orbit.
2. Coloured noise is generated by filtering white noise with the filter described in Equation 13. The resulting coloured noise is normalised to match the combined standard deviation of observation and prediction in Table 5. Then, the mean of both observation and prediction is added to the noise to obtain the desired rms error. The sum of the true orbit and the coloured noise is the predicted orbit.
3. A 10-order Chebyshev is fitted during each satellite pass on the predicted orbit. This is the ephemeris that will be used by the receiver to calculate the location of the satellite.

**Velocity Error**   To quantify the effect of the ephemeris error in a Doppler scenario, its velocity error is considered. Two approaches are possible for a receiver to obtain the satellite velocity: approximating from the satellite position, or calculating from a separately fitted velocity model. From simulation we found these two methods give similar rms error at around 60 mm/s. To save on the size of the ephemeris message, this paper uses the former method and approximates Chebyshev derivative method to calculate satellite velocity.

### 5.1.2 | Ephemeris Method 2

This method is more ideal and may not be achievable until advanced methods for ephemeris determination and prediction are available. This method arises primarily for comparison with other researches on lunar navigation. It is based on NASA's requirement of a lunar navigation satellite, which specifies the standard deviation of the ephemeris error to be 4.48 m in position and 0.4 mm/s in velocity (NASA, 2022). This is presumably for a higher orbit such as ELFO since the requirement specifies that the navigation satellite should be able to provide service to LLO. To capture the difficulty in accurately predicting LLO, this error is scaled linearly with respect to orbit prediction error using HALO, the lunar propagator in section 5.1.1. The resulting ephemeris error has standard deviation of 9.32 m and 1.8 mm/s, as detailed in Table 6.

**TABLE 6**
NASA required Ephemeris error and HALO prediction error of ELFO and LLO (Granier et al., 2024)(NASA, 2022)

|  | ELFO | LLO |
|---|---|---|
| Ephemeris position error (m) | 4.48 | $\frac{4.48}{24.66} \times 51.29 = 9.32$ |
| HALO position error (m) | 24.66 | 51.29 |
| Ephemeris velocity error (mm/s) | 0.4 | $\frac{0.4}{10.1} \times 45.8 = 1.8$ |
| HALO velocity error (mm/s) | 10.1 | 45.8 |

## 5.2 | Satellite Clock

The satellite clock also affects the location accuracy. (Bauer et al., 2016) shows that the clock stability of LRO is $2e-13 s/s$ after one orbital period. The error from the satellite clock is then

$$\sigma_{clk,sat} = c \cdot \sigma_{f,sat}$$

where $\sigma_{f,sat} = 2e-13 s/s$ is the deviation of the clock's fractional frequency, and $\sigma_{clk,sat} = 6e-5 m/s$.

## 5.3 | Receiver Clock

The receiver clock has a larger effect on position errors than the satellite clock as it is typically cheaper and lighter. Section II.5.b of Coimbra et al. (2024) details four different clocks that are suitable for lunar rovers. The SRS PRS 10 clock is used in this work because it has the best performance with less than 1 kg weight. Its stability $\sigma_{f,rec}$ is $8e-12 s/s$. The overall error from the receiver clock is (Coimbra et al., 2024)

$$\sigma_{clk,rec} = c \cdot \sigma_{f,rec}$$
$$\sigma^2_{f,rec} = \frac{h_0}{2T_s} + 4h_{-1} + \frac{8}{3}\pi^2 T_s h_{-2}$$

where $h_0 = 1.3e-22\ s^2/s$, $h_{-1} = 2.3e-26\ s^2/s^2$, $h_{-2} = 3.3e-31\ s/s^2$ are the power spectral density coefficients that describe the stability of the SRS PRS 10 clock, and $T_s$ is the sampling time.

## 5.4 | Measurement

The measurement noise due to carrier tracking is another contributor to location accuracy. Section II.5.a of Coimbra et al. (2024) details the expected standard deviation. It assumes the measurement noise is related to the received carrier-to-noise ratio $C/N_0$, which depends on the off-boresight angle. The off-boresight angle is the angle between the satellite antenna direction and the line of sight between the satellite and the receiver. The smaller the off-boresight angle, the stronger the received signal. However, the off-boresight angle can be as large as 65 deg for an LLO, which would generate low $C/N_0$ using the traditional high gain antenna, severely affecting the location accuracy. Therefore, it is assumed that an omnidirectional antenna with a constant gain will be used. This is realistic because the LRO is equipped with an omnidirectional antenna (Currier et al., 2018), and NASA's omnidirectional antenna has a relatively constant gain from 0 to 70 deg (Hilliard, 1987). The calculation is shown below with the symbols listed in Table 7.

$$C/N_0 = P_r + g/T - k$$

$$P_r = EIRP - 20 \log_{10}\left(\frac{4\pi f_0 \|r_{sr}\|}{c}\right)$$

$$g/T = G_r - T_{eq}$$

$$T_{eq} = 10 \log_{10}\left(T_{sys} + 290\left(10^{NF_{LNA}/10} - 1\right)\right)$$

$$\sigma_{meas} = \frac{c}{2\pi f_0 T}\sqrt{\frac{B_{PLL}}{C/N_0}\left(1 + \frac{1}{2TC/N_0}\right)}$$

**TABLE 7**
Symbols used in $C/N_0$ and $\sigma_{meas}$ calculations

| Symbol | Meaning | Value |
|---|---|---|
| $G_r$ | Gain of receiver antenna (dB) | 22.0 |
| $T_{sys}$ | System noise temperature of receiver antenna (K) | 113 |
| $NF_{LNA}$ | Noise figure of receiver's LNA (dB) | 1 |
| $B_{PLL}$ | PLL bandwidth (Hz) | 10 |
| $T$ | Coherent time integration (s) | 0.02 |
| $EIRP$ | Equivalent isotropic radiated power (dBi) | 0 |
| $k$ | Boltzmann constant (dBW/(KHz)) | -228.6 |
| $g/T$ | Gain-to-noise-temperature ratio | |

## 5.5 | Total measurement error

Taking into account the previous four factors, the total measurement error can be written as

$$\sigma_{tot}^2 = \sigma_{vel}^2 + \sigma_{clk,sat}^2 + \sigma_{clk,rec}^2 + \sigma_{meas}^2, \tag{14}$$

where the errors have units $km/s$. This error will be used for DOP calculation in Section 7.

# 6 | SIMULATION

## 6.1 | Setup

Table 8 shows some performance of the proposed orbit.

Figure 5 illustrates the satellite trajectory during one satellite pass. The signal availability time is roughly 12 minutes per orbital period, depending on the location of the receiver.

In the following sections, 100 Monte-Carlo simulations are performed for each scenario. The epoch time is randomly selected within Mar 20, 2024 and June 10, 2024. The orbit is assumed to have an initial mean anomaly of 180 deg, so that it starts near the South Pole. The receiver position is randomly distributed with a uniform distribution between

TABLE 8
Orbit performance

| Parameter | Value |
|---|---|
| Orbital period (h) | 2 |
| Mean signal availability time (min) | 12 |
| Coverage in latitude (°) | 70 - 90 |
| Mask angle (°) | 5 |
| Maximum off-boresight angle (°) | 65 |

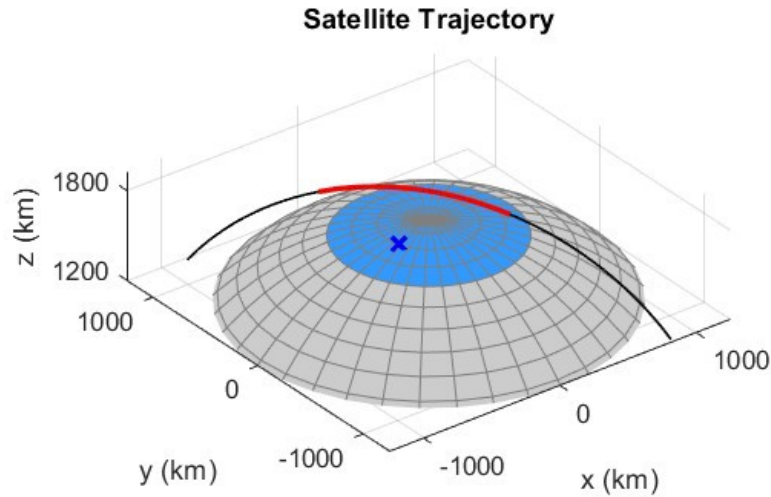

**FIGURE 5** Satellite trajectory over one pass. Blue surface: North Pole region (70° − 90° latitude); blue cross: receiver; red line: satellite visible to receiver; black line: satellite not visible to receiver.

- Latitude from 70 to 90 deg;
- Longitude from 0 to 360 deg;
- Altitude from -10 to +10 km from lunar surface.

The measured Doppler is calculated by Equation 1 when the elevation angle is more than the mask angle at 5 deg. When the elevation angle is less than 5 deg, signal is assumed to be unavailable.

## 6.2 | Location Accuracy

The location accuracy is examined using both ephemeris methods outlined in subsection 5.1. It considers the case with one satellite pass, and the case where multiple consecutive satellite passes are available. Two types of errors are identified: 99% error is the 99th percentile out of the 100 simulations, and the mean error is the average error out of the same 100 simulations.

### 6.2.1 | One satellite Pass

This subsection shows the algorithm performance when only one satellite pass is available. It analyses the accuracy in each step. It also finds the empirical mirror identification rate based on the value of the cost function.

**Ephemeris 1**

Figure 6(a) shows the histogram of position errors resulting from each of the three steps using ephemeris 1. It can be seen that the error reduces significantly from step 1 to 3, illustrating the effectiveness of each step.

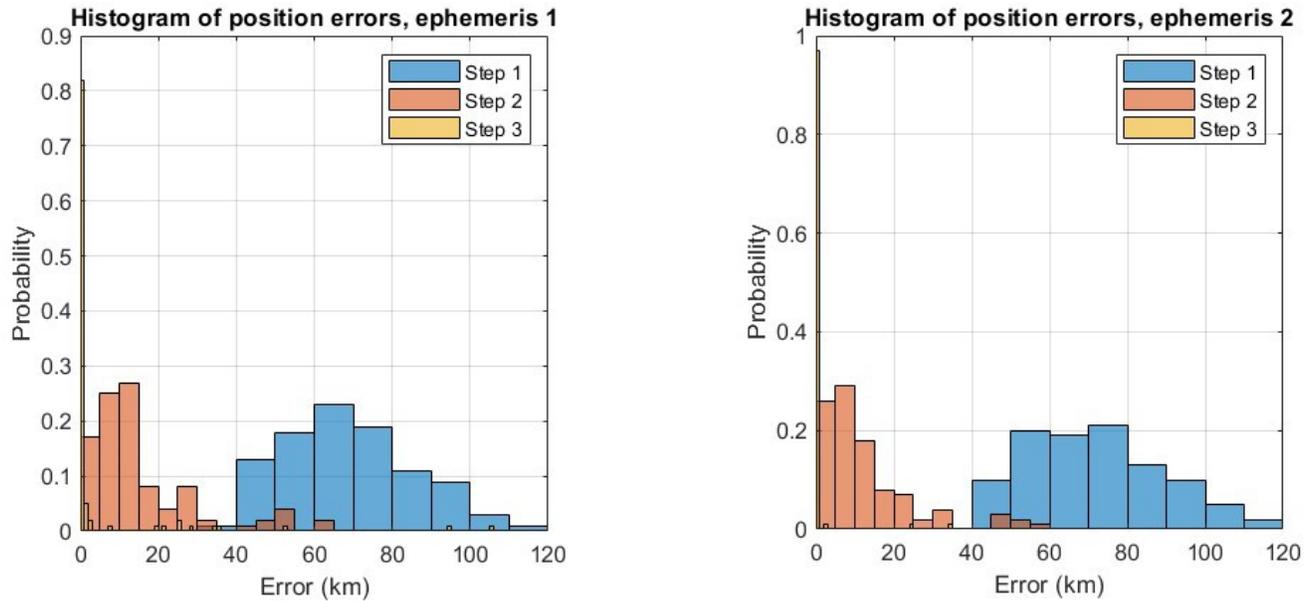

(a) Ephemeris 1

(b) Ephemeris 2

**FIGURE 6** Histograms of position error in each step with both ephemeris methods

### Ephemeris 2

Figure 6(b) shows the histogram of position errors using ephemeris 2. Compared with ephemeris 1, the step 1 and 2 accuracy are similar, while the step 3 is slightly better. This is because steps 1 and 2 are not meant for high accuracy, rather they are to increase the convergence rate and speed up the algorithm. The reason for only a slightly better result in step 3 is because the bin size is 1 km, and both ephemeris result in estimations better than 1 km most of the time. Table 9 summarises the position estimation performance with only one satellite pass.

**TABLE 9**
Mean position error in each step using both ephemeris methods

| Ephemeris | Step | Mean error (km) |
|---|---|---|
| 1 | 1 | 64.6 |
|   | 2 | 10.6 |
|   | 3 | 0.1 |
| 2 | 1 | 70.2 |
|   | 2 | 8.6 |
|   | 3 | 0.1 |

### Number of iterations

The number of iterations in step 2 and 3 are shown in Figure 7. In a few cases 100 iterations are performed, which hit the upper limit of the algorithm and are omitted in the histograms.

For ephemeris 1, there is a 1% probability of having 100 iterations for step 2, and a further 4% chance of having 100 iterations for step 3. This is due to the ephemeris error being too large, and the algorithm starts oscillating between several points, resulting in too many iterations until the local minimum point. This issue does not occur for ephemeris 2, possibly due to the cost function being more well-behaved as it has a smaller ephemeris error. For ephemeris 1, the average iterations to complete both steps 2 and 3 are 18.76; for ephemeris 2, only 13.82 iterations are required.

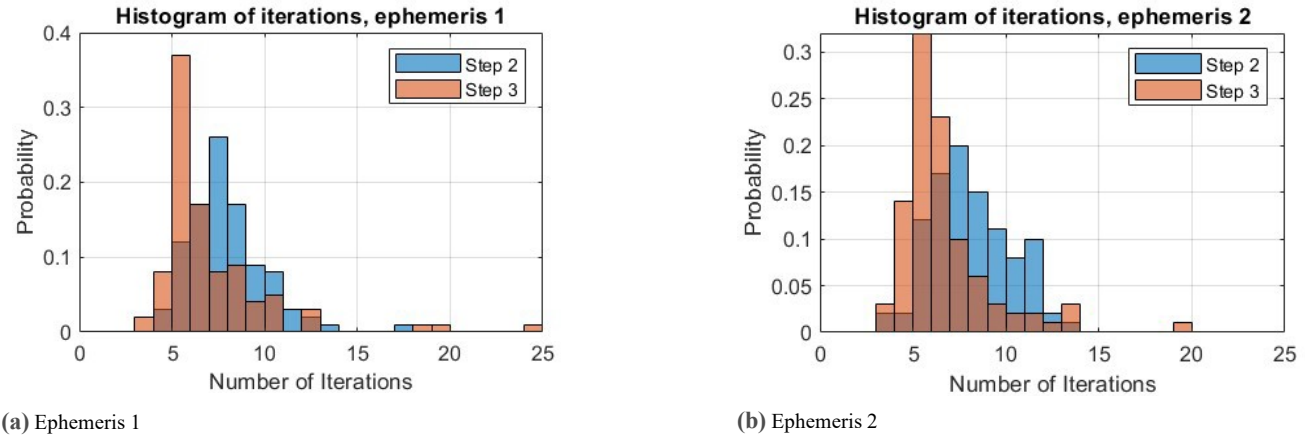

**(a)** Ephemeris 1

**(b)** Ephemeris 2

**FIGURE 7** Histograms of number of iterations in step 2 and 3 with both ephemeris methods

**Mirror identification rate**

Using ephemeris 1, the mirror location is found by using the simpile method of reflecting the location estimate about the satellite subtrack, and applying step 3 to find the local minimum cost function. Then, the cost function values of the two locations are compared, and the true location is the one with a lower cost function. With 100 simulations, the mirror identification rate is at 53%. This low success rate shows that the cost function method is not suitable for mirror identification due to the large noise in the measurements. Using multiple satellite passes is a much more reliable method, with 100% success rate.

### 6.2.2 | Multiple Satellite Passes

This subsection shows the algorithm performance when multiple satellite passes are available. The mirror location is identified using the methods in subsection 4.4, which always identifies the correct receiver location.

Figure 8 shows the position error using ephemeris 1 and 2 when 10 consecutive satellite passes are available. It shows the 99% and the mean error for each ephemeris. Table 10 records some important results from this plot. It can be seen that the largest improvement is from 1 pass to 2 passes. It should be noted that the mirror location requires 2 satellite passes to be correctly identified, making 2 passes a balanced choice between estimation time and position accuracy.

**TABLE 10**
99% and mean error using 1, 2, and 10 satellite passes using both ephemeris methods

| Ephemeris | Error type | Position error (m) | | |
|---|---|---|---|---|
| | | 1 pass | 2 passes | 10 passes |
| 1 | 99% | 21,000 | 2,200 | 490 |
| | Mean | 790 | 270 | 110 |
| 2 | 99% | 670 | 75 | 6.1 |
| | Mean | 36 | 10 | 1.9 |

**Comparison with ELFO**

Coimbra et al., 2024 uses an ELFO for single satellite navigation. They also use the Doppler method, and assume that the ephemeris meets the NASA requirements. Its results are compared with ephemeris method 2 using LLO proposed in this thesis in Table 11. It can be seen that the time to get sub-10m 99% error is slightly reduced over the ELFO case, but the time to get sub-10m mean error is greatly reduced from 12 hours to 4 hours. This is because the LLO accuracy depends heavily on the geometry of the receiver. The 99% error is almost entirely due to the high GDOP areas near the edge of the coverage area shown in Figure 10. On the other hand, the mean error is less affected by the geometry and shows the expected time to reach sub-10m accuracy. This shows that LLO provides a better accuracy compared to ELFO, largely due to the fast dynamics of the satellite with respect to the receiver.

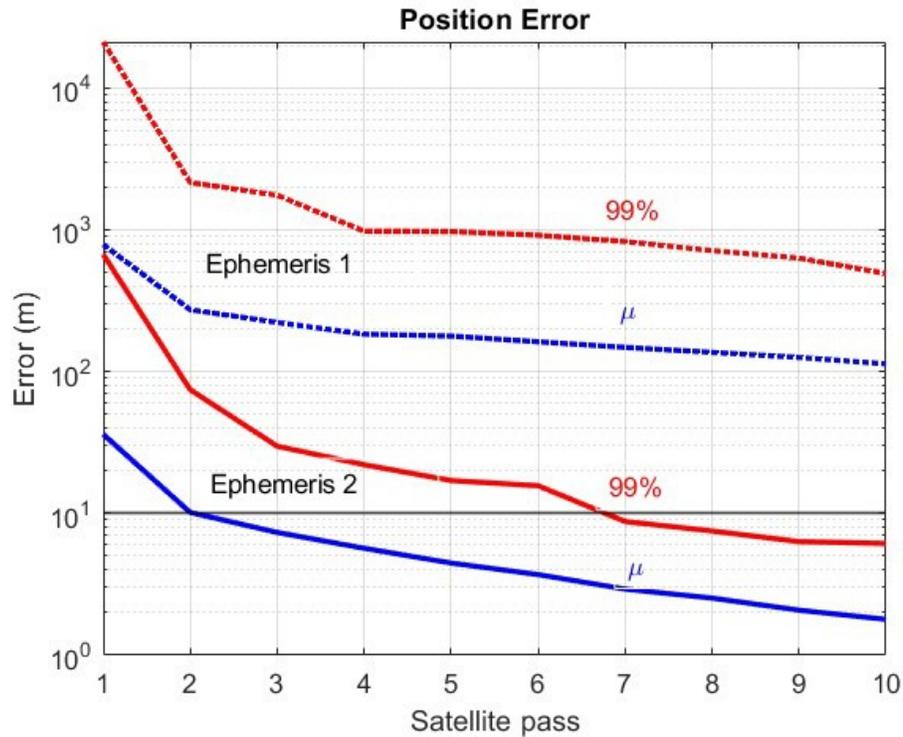

**FIGURE 8** 99% and mean position error using ephemeris methods 1 and 2. Horizontal line is 10 m.

**TABLE 11**
Comparison of ELFO and LLO on time to reach sub-10m accuracy Coimbra et al., 2024

| Error type | Time to reach sub-10m accuracy (h) | |
|---|---|---|
| | ELFO | LLO |
| 99% | 16 | 14 |
| Mean | 12 | 4 |

## 6.3 | Error Budget

The individual contribution of each error from the error budget in subsection 5 is investigated. Only the ephemeris 2 case is considered as other errors should have the same effect with either ephemeris 1 or 2. Figure 9 shows the 99% and mean error resulting from each of the error source. Table 12 summarises the individual contribution of each error source to the mean error after 10 satellite passes. It can be seen that ephemeris causes 88% of total error, followed by receiver clock at 10%. The carrier tracking error and satellite clock error is negligible compared to ephemeris and receiver clock. This shows that the orbit accuracy is highly dependent on the quality of the ephemeris, which explains why there is a huge performance difference between ephemeris 1 and 2.

## 7 | DILUTION OF PRECISION

Dilution of Precision (DOP) is used to analyse the expected accuracy of the location estimate as a consequence of the particular geometry used in the solution (Psiaki, 2021). A lower DOP value means a better estimation accuracy. It depends only on the ephemeris and the receiver location and gives an indication of how good the location estimate is. Consider the true location $x$ and

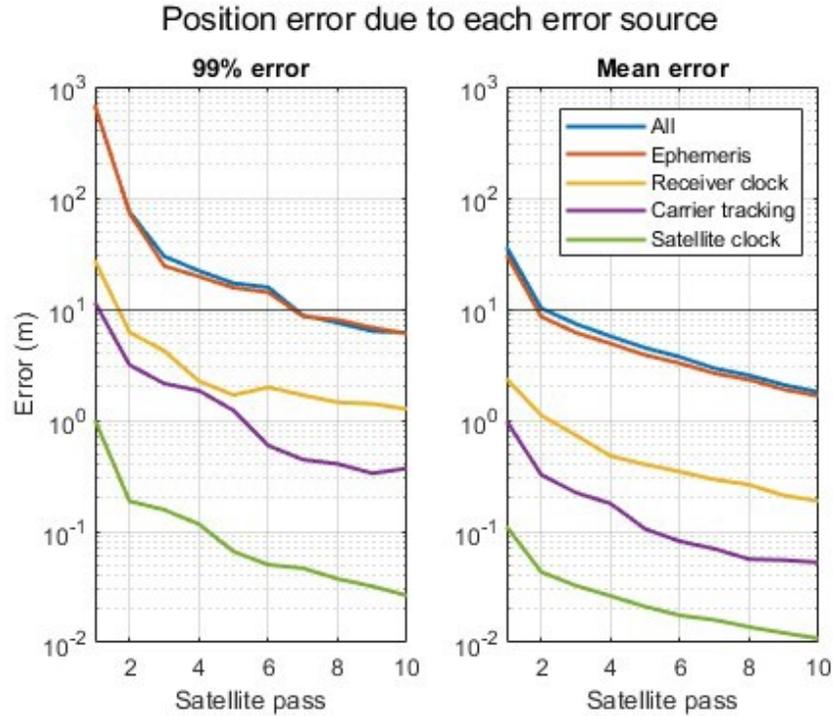

**FIGURE 9** Position error due to each error source using ephemeris 2

**TABLE 12**
Contribution of each error source

| Error Source | Error (m) | Percentage (%) |
|---|---|---|
| Ephemeris | 1.65 | 87 |
| Receiver Clock | 0.19 | 10 |
| Carrier Tracking | 0.05 | 3 |
| Satellite Clock | 0.01 | 1 |
| Total | 1.90 | 100 |

estimated location $\hat{x}$. They are related by error $\Delta x$ by

$$x = \hat{x} + \Delta x$$
$$f(x) = f(\hat{x} + \Delta x)$$
$$f(x) = f(\hat{x}) + \frac{\partial f(\hat{x})}{\partial x^T}\Delta x + \ldots$$
$$f(x) - f(\hat{x}) = H(\hat{x})\Delta x$$
$$\lambda_0 \sigma_{tot} = H(\hat{x})\Delta x$$
$$\Delta x = H^{-1}(\hat{x})(\lambda_0 \sigma_{tot})$$

Assuming $\Delta x$ and $\sigma_{tot}$ are zero-mean, taking the variance on both sides gives

$$\begin{matrix} var(x) & cov(x,y) & cov(x,z) \\ cov(x,y) & var(y) & cov(y,z) \\ cov(x,z) & cov(y,z) & var(z) \end{matrix} = \left( H_{DOP}^T H_{DOP} \right)^{-1} (\hat{x})$$

where $H_{DOP} = \lambda_0 H \sigma^{-1}_{tot}$. Therefore, the Geometric DOP (GDOP) is

$$GDOP = \sqrt{var(x) + var(y) + var(z)} = \sqrt{trace\left(\left(H^T_{DOP} H_{DOP}\right)^{-1}\right)}. \quad (15)$$

Figure 10 shows a GDOP map for 10 consecutive satellite passes. It is a top-down view from the North Pole, covering latitudes from 70 to 90 deg. It separates the North Pole region into grids of 1 deg in latitudes and 5 deg in longitudes. All points are on lunar surface. A lower GDOP value indicates a better accuracy of the location estimate. From the graph, the GDOP is lower along the satellite subtrack, and is worse further away from the satellite subtrack. This implies that the viewing angle from receiver to satellite could be dominant in GDOP. This is similar to the GPS case, where the vertical DOP would be bad if there are no satellites directly above the receiver. Due to the slow rotation of the Moon, consecutive satellite passes do not cause a lot of differences in GDOP values.

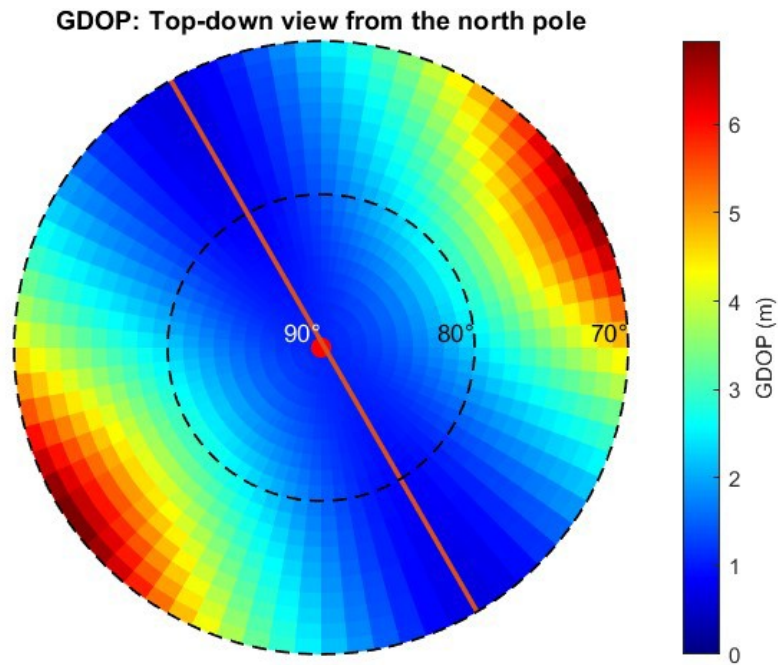

**F I G U R E 10** Top-down view of GDOP values after ten satellite passes. Latitudes are from 70 to 90 deg. Red point at the centre is the North Pole. Red line shows the 5th satellite subtrack.

## 8 | CONCLUSION

In conclusion, this work proposes a single satellite navigation system at lunar North Pole for a stationary receiver. It utilises the Doppler frequency profile received by the receiver and maps it to a single location with sub-10m accuracy without any prior knowledge of the receiver position.

The LLO satellite has a period of 2 hours and provides approximately 12 minutes of signal available time during each orbital period. It covers the lunar North Pole region where the latitude is more than 70 deg. The location algorithm consists of three steps: an algebraic solution, a constrained non-linear least square problem, and an unconstrained non-linear least square problem. The use of this three-step algorithm greatly increases the convergence rate when no prior knowledge of the receiver position is available. It also eliminates the mirror location with at least two satellite passes. Two ephemeris methods are used, the first one can be achieved today, while the second one requires specialised techniques to meet the NASA requirements. The 99% position accuracy is found to be sub-500m when using ephemeris 1, and sub-10m when using ephemeris 2. The time taken to reach the same level of accuracy is less compared to other orbits such as ELFO. An error budget is suggested and individual impact of each error source is simulated. Using ephemeris 2, it is found that the ephemeris contributes almost 90% of the final position error, while receiver clock contributes 10%. Carrier tracking and satellite clock errors have negligible effect. DOP is used as a metric for the location accuracy, and is found to be highly correlated with elevation angle. The DOP, and thus the accuracy, is better with a large elevation angle.

Future work could include further refining the ephemeris error and the total measurement error for this navigation system, and applying sequential filters instead of batch filters to minimise memory storage requirements. It is also desirable to include the receiver clock offset in the estimated states. Finally, the receiver can be allowed to have non-zero velocity.

## ACKNOWLEDGMENTS

The author would like to thank Ms. Kaila M. Y. Coimbra (Stanford University) and Mr. Quentin Granier (University of New South Wales) for their help and support in this work.